\documentclass[10pt, conference]{IEEEtran}
\IEEEoverridecommandlockouts
\usepackage{color}
\usepackage{array}
\usepackage{listings}
\usepackage{xspace}
\usepackage{amsmath}
\usepackage{tikz}
\usepackage{url}
\usepackage{multirow}
\usepackage{stmaryrd}
\usepackage{etoolbox}
\usepackage[misc]{ifsym}
\usepackage{balance}

\newcommand{\code}[1]{\texttt{\footnotesize #1}}

\newcommand{\figref}[1]{Fig.~\ref{#1}}
\newcommand{\tabref}[1]{Table~\ref{#1}}

\newcommand{\secref}[1]{Section~\ref{#1}}
\newcommand{\tblref}[1]{Table~\ref{#1}}
\newcommand{\smalltitle}[1]{{\smallskip \noindent \bf  {#1}.\ }}

\newtoggle{longversion}
\toggletrue{longversion}

\newcommand{\approach}{ACS\xspace}

\begin{document}
%
\title{Precise Condition Synthesis for Program Repair}


%


%

\author{\IEEEauthorblockN{Yingfei
    Xiong\IEEEauthorrefmark{1}\IEEEauthorrefmark{2}, Jie
    Wang\IEEEauthorrefmark{1}\IEEEauthorrefmark{2}, Runfa
    Yan\IEEEauthorrefmark{3}, Jiachen
    Zhang\IEEEauthorrefmark{1}\IEEEauthorrefmark{2}, Shi
    Han\IEEEauthorrefmark{4}, Gang Huang\textsuperscript{(\Letter)}\IEEEauthorrefmark{1}\IEEEauthorrefmark{2}, Lu
  Zhang\IEEEauthorrefmark{1}\IEEEauthorrefmark{2}\thanks{We
    acknowledge participates of Dagstuhl 17022, especially Julia Lawall, Yuriy
    Brun, Aditya Kanade, and Martin Monperrrus, and anonymous
    reviewers for their comments on the paper, and Xuan-Bach D.
    Le, David Lo, and Claire Le Goues for discussion on their experiment data. This work is
    supported by the National Basic Research Program of China under
    Grant No. 2014CB347701, the National Natural Science Foundation of
    China under Grant No. 61421091, 61225007, 61672045, and Microsoft Research Asia Collaborative Research Program 2015, project ID FY15-RES-OPP-025.}}
\IEEEauthorblockA{\IEEEauthorrefmark{1}Key Laboratory of High Confidence Software
  Technologies (Peking University), MoE, Beijing 100871, China}
\IEEEauthorblockA{\IEEEauthorrefmark{2}EECS, Peking University,
Beijing 100871, China, \{xiongyf, 1401214303, 1300012792, hg, zhanglucs\}@pku.edu.cn}
\IEEEauthorblockA{\IEEEauthorrefmark{3}SISE, University of Electronic Science and
  Technology of China, Chengdu 610054, China, yanrunfa@outlook.com}
\IEEEauthorblockA{\IEEEauthorrefmark{4}Microsoft Research,
  Beijing 100080, China, shihan@microsoft.com}
}


\maketitle

\begin{abstract}
  Due to the difficulty of repairing defect, many research efforts
have been devoted into automatic defect repair. Given a buggy program
that fails some test cases, a typical automatic repair technique tries
to modify the program to make all tests pass. However, since the test
suites in real world projects are usually insufficient, aiming at
passing the test suites often leads to incorrect patches. This problem
is known as weak test suites or overfitting.

  In this paper we aim to produce precise patches, that is, any patch
we produce has a relatively high probability to be correct. More
concretely, we focus on condition synthesis, which was shown to be
able to repair more than half of the defects in existing approaches.
Our key insight is threefold. First, it is important to know what
variables in a local context should be used in an ``if'' condition,
and we propose a sorting method based on the dependency relations
between variables. Second, we observe that the API document can be
used to guide the repair process, and propose document analysis
technique to further filter the variables. Third, it is important to
know what predicates should be performed on the set of variables, and
we propose to mine a set of frequently used predicates in similar
contexts from existing projects.

Based on the insight, we develop a novel program repair system,
\approach, that could generate precise conditions at faulty locations.
Furthermore, given the generated conditions are very precise, we can
perform a repair operation that is previously deemed to be too
overfitting: directly returning the test oracle to repair the defect.
Using our approach, we successfully repaired 17 defects on four
projects of Defects4J, which is the largest number of fully
automatically repaired defects reported on the dataset so far. More
importantly, the precision of our approach in the evaluation is
73.9\%, which is significantly higher than previous approaches, which
are usually less than 40\%.

\end{abstract}


%
\IEEEpeerreviewmaketitle
\section{Introduction}

\smalltitle{Motivation} Given the difficulty of fixing defects, recently a lot of research efforts
have been devoted into automatic program repair
techniques~\cite{GenProg,PAR,Autofix-E,SemFix,DirectFix,staged,RSRepair,r2fix,MintHint,DBLP:journals/chinaf/QiMWDG12}.
Most techniques generate a patch for a defect aiming at satisfying a
specification. A frequently used specification is a test suite. Given a
test suite and a program that fails to pass some tests in the test
suite, a typical repair technique modifies the
program until the program passes all tests.

However, the tests in real world projects are usually
insufficient, and passing all tests does not necessarily mean that the
program is correct. A patch that passes all tests is known as a
\emph{plausible} patch~\cite{qi15}, and we use \emph{precision} to
refer the proportion of defects correctly fixed by the first plausible patch
among all plausibly fixed defects. 
We argue that
precision is a key quality attribute of program repair systems.  If
the precision of a
repair system is similar to or higher than human developers, we can trust the patch generated
by the system and deploy them immediately. On the other
hand, if the precision of a repair system is low, the developers
still need to manually review the patches, and it is not clear whether
this process is easier than manual fix. As an
existing study~\cite{Tao:2014:AGP:2635868.2635873} shows, if the
developers are provided with low-quality patches, the performance of
the developers is even lower than those who are provided with no
patches.

However, the precisions of existing testing-based repair techniques are not
high~\cite{qi15,smith2015cure,martinez2015automatic}. As studied by Qi
et al.~\cite{qi15}, GenProg, one of the most well-known program repair
techniques, produced plausible patches for 55 defects, but only two
were correct, giving a precision of 4\%. 

The reason for the low precision, as studied by Long et
al.~\cite{long2016analysis}, is that correct patches are sparse in the
spaces of repair systems, while plausible ones are
relatively much more abundant. In their experiments, often hundreds
or thousands of plausible patches exist for a defect, among which only
one or two are correct. It is difficult for the repair system to
identify a correct patch from the large number of plausible patches.

A fundamental way to address this problem is to rank the patches by their
probabilities of being correct, and return the
plausible patch with the highest probability, or report failure when
the highest probability is not satisfactory. This is known as \emph{preference bias} in inductive
program synthesis~\cite{kitzelmann2009inductive}. 
Research efforts have been made toward this direction.
Prophet~\cite{Long2016} and HistoricalFix~\cite{Xuan2016History} rank the patches using models learned from
existing patches. DirectFix~\cite{DirectFix}, Angelix~\cite{Mechtaev} and Qlose~\cite{dqlose} rank
the patches by their distance from the original program.
MintHint~\cite{MintHint} ranks the patches by their statistical correlation with the expected results.
However, the
precisions of these approaches are not yet satisfactory. For example,
Prophet~\cite{Long2016} and Angelix~\cite{Mechtaev} have precisions of
38.5\% and 35.7\% on the GenProg
Benchmark~\cite{EightDollar}. 


An important reason for this imprecision, as we conjecture, is that the
existing ranking approaches are too coarse-grained. As will be shown
later, existing approaches cannot 
distinguish many common plausible ``if'' conditions from the correct
condition, and will give them the same rank.

To overcome this problem, in this paper we aim to provide more
fine-grained ranking criteria for condition synthesis.
Condition synthesis tries to insert or modify an ``if'' condition to
repair a defect. Condition synthesis has been used in several existing
repair systems~\cite{staged,Mechtaev,xuan2016nopol} and is shown to be
one of the most effective techniques. For example, among all defects
correctly repaired by SPR~\cite{staged}, more than half of them are
fixed by condition synthesis.

Our approach combines three heuristic ranking techniques that exploit
the structure of the buggy program, the document of the buggy program,
and the conditional expressions in existing projects. More concretely, we
view the condition synthesis process as two steps. (1)
\emph{variable selection:} deciding what variables should be used in
the conditional expression, and (2) \emph{predicate selection:} deciding
what predicate should be performed on the variables. For example, to
synthesize a condition \code{if(a>10)}, we need to first select the
variable ``\code{a}'' and then select the predicate ``\code{>10}''. Based
on this decomposition, we propose the following three techniques for ranking variables and predicates, respectively.
\begin{itemize}
\item \emph{Dependency-based ordering.} We observe that the principle
  of locality holds on variable uses: the more recent a variable in a
  topological ordering of dependency is, more likely it will be used
  in a conditional expression. We use this order to rank variables.
\item \emph{Document analysis.} We analyze the javadoc comments
  embedded in the source code. When variables are mentioned for a
  particular class of conditional expressions, we restrict the
  selection to only the mentioned variables when repairing
  the conditional expressions of the class.
\item \emph{Predicate mining.} We mine predicates from
  existing projects, and sort them based on their frequencies in
  contexts similar to the target condition.
\end{itemize}

To our knowledge, the three techniques are all new to program repair.
We are the first to propose the locality of variable uses and apply it
to program repair. We also are the first to utilize the documentation
of programs to increase precision. Finally, predicate mining is the
first technique that automatically mines components of patches from
the source code (in contrast to those mining from
patches~\cite{Long2016,Xuan2016History,genesis}) of existing projects.

As will be shown later, the combination of the three techniques gives
us high precision in program repair. Based on the high precision, we
further employ a new repair method that was considered to be too
overfitting: directly returning the test oracle of the failed test to repair a faulty
method. 
This is considered overfitting because a test oracle is
usually designed for a specific test input, and it is not clear
whether this test oracle can be and should be generalized to other
inputs.
However, this overfitting is likely not to exist when the test input
belongs to a boundary case. A boundary case is a case that cannot be
captured by the main program logic, and for such a case usually a value
or an exception is directly returned. For example, a patch we
generated in our experiment, \code{if (marker==null) return false},
directly return the oracle value \code{false} when the input is \code{null}.
Since our condition synthesis is precise, if we can successfully synthesize a condition that
checks for a boundary case, it is probably safe to directly return the
oracle.

We have implemented our approach as a Java program repair system,
\approach (standing for Accurate Condition Synthesis), and evaluated
\approach on four projects from the Defects4J benchmark.
\approach successfully generated 23 patches, where 17 are correct,
giving a precision of 73.9\% and a recall of 7.6\%. Both the precision
and the recall
are the best results reported on Defects4J so far within our
knowledge.  Most importantly, the precision is significantly higher
than previous testing-based approaches,
which are usually less than 40\%. Furthermore, we evaluated the three
ranking techniques on the top five
projects from GitHub besides the four projects. The result suggests
that our approach may achieve a similar precision across a wide range
of projects.


\section{Motivating Example}\label{sec:overview}

\begin{figure}[ht]
  \centering
  \vspace{-0.3cm}
\begin{lstlisting}
 int lcm=Math.abs(mulAndCheck(a/gdc(a,b),b));
+if (lcm == Integer.MIN_VALUE) {
+  throw new ArithmeticException();
+}
 return lcm;
\end{lstlisting}
  \vspace{-0.3cm}
\caption{\label{fig:example1} Motivating Example}
\end{figure}

\smalltitle{Example} Figure~\ref{fig:example1} shows the Math99
defect in the Defects4J~\cite{just2014defects4j} benchmark fixed by \approach. This method calculates the least common
multiple and is expected to return a
non-negative value, but it fails to handle the case 
where the method \code{abs}
returns \code{Integer.MIN\_VALUE} at input
\code{Integer.MIN\_VALUE} due to the imbalance between
positives and negatives in integer representation
. Two tests cover this piece of code. Test 1 has input
\code{a=1}, \code{b=50}, and expects \code{lcm=50}. Test 2 has
input \code{a=Integer.MIN\_VALUE}, \code{b=1}, and expects
\code{ArithmeticException}. Test 1 passes and test 2
fails.

To fix the defect, \approach tries to directly return the test oracle,
and adds lines 2-4 to the program. In particular, \approach synthesizes
the condition at line 2. Since we
have only two tests, a condition is plausible if it evaluates to
\code{false} on test 1 and evaluates to \code{true} on test 2. Thus, there exists many plausible conditions that can make the
two tests pass, such as \code{b==1} or \code{lcm != 50}. As a result,
we need to select the correct condition from the large
space of plausible conditions.

Existing approaches are not good at this kind of fine-grained ranking.
For example, Prophet~\cite{Long2016} always assigns the same priority to \code{lcm!=50} and
\code{lcm==Integer.MIN\_VALUE} because, due to efficiency reasons,
Prophet can only consider variables in a condition. For another
example, Qlose~\cite{dqlose} always assigns the same priority to \code{b==1},
\code{lcm!=50}, and \code{lcm==Integer.MIN\_VALUE} because the three
conditions exhibit the same behavior in testing executions, which
Qlose uses to rank different patches.

To implement fine-grained ranking, \approach decomposes condition ranking
into variable ranking and predicate ranking, and using three
techniques to rank variables and predicates.

\smalltitle{Dependency-based ordering} Our observation is that the
principle of locality also holds on variable uses. If variable $x$ is
assigned from an expression depending on $y$, i.e., $x$ depends on
$y$, $x$ has a higher chance than $y$ to be used in a following conditional
expression. The intuition is
that $y$ is more likely to be a temporary variable whose purpose is to
compute the value of $x$. In our example, variable
\code{lcm} depends on variables \code{a} and \code{b}, and thus
\code{lcm} is more likely to be used in the ``if'' condition. Based on
the dependency relations, we can perform a topological sort of the
variables, and try to use most dependent variables in the
synthesis first.



\smalltitle{Document Analysis} We further observe that many Java
methods come with javadoc comments, documenting the functionality of
the method. If we extract information from the
javadoc comment, we could further assist condition synthesis. As the
first step of exploiting comments, in this paper we consider one type
of javadoc tags: \code{@throws} tag. Tag \code{@throws} describes an exception that
may be thrown from the method, such as the following one from Math73
in Defects4J~\cite{just2014defects4j}.
\begin{lstlisting}[basicstyle=\scriptsize\rmfamily,numbers=none,xleftmargin=0em]
/** ...
 * @throws IllegalArgumentException if initial is not between
 * min and max (even if it <em>is</em> a root)
**/
\end{lstlisting}
According to this comment, an exception should be thrown when
the value of \code{initial} is outside of a range.
To exploit this information for variable selection, we analyze the
subject of the sentence. If the subject mentions any variable in the
method, we associate the variable with the exception.
When we try to generate a guard condition for such an exception, we
consider only the variables mentioned in the document. Since
programmers may not refer to the variable name precisely, we use fuzzy
matching: we separate
variable names by capitalization, i.e., \code{elitismRate} becomes
``elitism'' and ``rate'', and determine that a variable is mentioned
in the document
if the last word (usually the center word) is mentioned.

Note that our current approach only makes the above lightweight use of javadoc
comments, but more sophisticated document analysis techniques may be used
to obtain more information, or even directly generate the condition~\cite{zhai2016automatic}. This
is future work to be explored.



\smalltitle{Predicate Mining} After we have an ordered list of
variables, we select predicates for the variables one by one. Based on
our observations, we have the insight that the predicates used in
conditional expressions have highly sparse and skewed conditional
distribution given the contexts of the expressions. 
We currently use variable type, variable name, and/or the surrounding
method name as context. 
For example, given
an integer variable \code{hour}, predicates such as \code{<=12} or
\code{>24} are often used. In our running example, \code{lcm}
indicates the least common multiple, on which
\code{==Integer.MIN\_VALUE} is more frequently used than a large
number of observed alternatives such as \code{!= 50}. 
For another example, in methods whose names contain
``factorial'', predicates such as \code{<21} is often used, because $20!$
is the largest factorial that a 64bit integer can represent.

Based on such insights, we prioritize and prune predicates based on their conditional
distributions of surrounding contexts. We approximate such conditional
distributions based on the statistics against a large scale repository
of existing projects. Concretely, we search predicates under 
similar contexts in a large repository, and rank them by their occurrences.

\smallskip
Combining the three techniques, we could
successfully synthesize \code{lcm==Integer.MIN\_VALUE}
at line 2. Since this condition is only valid for one value of
\code{lcm}, it is likely to be a boundary case and thus we can safely
generate the patch.



\section{Approach}\label{sec:approach}

In this section we explain the details of our approach. The input of
our approach consists of a program, a test failed by the program, a
set of tests passed by the program. The output is a patch on the
program.

\subsection{Overview}
We use two types of templates to fix defects. The first type is to
directly return the oracle as mentioned in the introduction. We first
identify the last executed statement $s$ in the failed test, and then insert
one of the following statement before $s$ to prevent the failure.
\begin{itemize}
\item \emph{Value-Returning}. \code{if ($c$) return $v$;}
\item \emph{Oracle-Throwing}. \code{if ($c$) throw $e$;}
\end{itemize}
When the failed test expects a return value, value-returning is used,
otherwise oracle-throwing is used. Here $v$ or $e$ is the expected
return value or exception, and $c$ is the synthesized condition. We
also use heuristic rules to check whether the synthesized $c$ is a boundary
check, and discard the patch if it 
is not a boundary check.

We always
insert before the last executed statement because, if the defect leads
to crash, for example, the program fails to check a null pointer, we
usually need to place the guarded return statement right before the
crashed statement. 

The second type is the modification of an existing condition. We first
locate a potentially faulty ``if'' condition $c'$, and then apply one of the
following modifications based on the result of predicate switching.
\begin{itemize}
\item \emph{Widening}. \code{if ($c'$) $\Rightarrow$ if ($c'$ || $c$)}
\item \emph{Narrowing}. \code{if ($c'$) $\Rightarrow$ if ($c'$ \&\& !$c$)}
\end{itemize}

We locate the potentially faulty condition by combining spectrum-based fault
localization (SBFL)~\cite{Jones:2002:VTI:581339.581397} and predicate
switching~\cite{zhang2006locating}.
Both approaches be used to detect potentially faulty conditions.
SBFL scales better while predicate switching
is more precise. In our approach, we first use
SBFL to localize a list of potentially
faulty methods, and then use predicate switching on demand within each
method. In this way we achieve a balance between precision and
scalability. The SBFL formula we used in our implementation is 
Ochiai~\cite{abreu2007accuracy}, which is shown to be among the
most effective spectrum-based fault localization algorithms~\cite{Xuan:2014:LCM:2705615.2706097,
  steimann2013threats,Xie:2013:TAR:2522920.2522924}.

If predicate switching negates the original condition from \code{true}
to \code{false}, we apply the narrowing template, otherwise we apply
the widening template. Here $c$ is the synthesized condition.

In both types of templates, we need to synthesize condition $c$. We
require $c$ to capture the failed test execution,
i.e., evaluating to \code{true} at the failed test execution. We first produce a ranked
list of variables to be used in the condition. Then for each
variable $x$, we produce a sorted list of predicates to be applied on the
variable. For each predicate $p$, we validate
whether they can form a condition $p(x)$ that capture the failed test
execution, i.e., evaluates to \code{true} on target condition
evaluation. If so, we synthesize a condition $p(x)$ and run all tests
to validate the plausibility of the patch. In this
paper we only consider synthesizing conditions containing one
variable, but note that our idea is general and may be extended to
conditions containing multiple variables. 

We always apply the oracle-returning templates first and the
condition-modification templates second. We return the first plausible
patch if it is found within the time limit, otherwise we report a failure.

Some of the above steps require more detailed explanations, and we
explain them one by one in the rest of the section.

\subsection{Returning the Oracle}
\smalltitle{Extracting the Oracle}
When applying the oracle returning template, we need to copy the test
oracle from the test code to the body of the
generated conditional statement. There are several different cases. (1) The
oracle is a constant. In this case we only need to directly copy the
text of the constant. (2) The oracle is specified via
\code{(expected=XXXException.class)} annotation. In this case we
throws an instance of \code{XXXException} by calling
its default constructor. (3) The oracle is a function mapping the test
input to the output.
This case is more complex. For example, the following piece of
code is a test method for Math3 defect in
Defects4J~\cite{just2014defects4j}.
\begin{lstlisting}
public void testArray() {
  final double[] a = { 1.23456789 };
  final double[] b = { 98765432.1 };
  Assert.assertEquals(a[0] * b[0],
    MathArrays.linearCombination(a, b), 0d);
}
\end{lstlisting}
Here the oracle is \code{a[0]*b[0]}, which is a function mapping input
$a$ and $b$ to the expected result. \approach inserts the following
statement into method \code{linearCombination} to fix the defect.
\begin{lstlisting}[numbers=none]
+ if (len == 1) {return a[0]*b[0];}
\end{lstlisting}
The condition \code{len==1} is generated using our condition synthesis
component and is not our concern here. The key is how to generate \code{a[0]*b[0]}.

To extract a functional oracle, we first identify the related pieces of code
by performing slicing. We first perform backward slicing from the
oracle expression (the oracle slice), then perform backward slicing
from the test input arguments (the input slices), and subtract the
input slices from the oracle slice. In this way we get all code
necessarily to be copied but not the code used to initialize the test
inputs. In this example, the oracle slice contains line 2, line 3 and
the expression \code{a[0]*b[0]}, the input slices contain line 2, line
3, and the expression \code{a} and \code{b}. By subtracting the latter
slices from the former, we get only the expression \code{a[0]*b[0]},
which should be copied to the generated ``if'' statement.
Finally, we rename the variables representing the test input arguments
(in this case, $a$ and $b$)
to the formal parameter names of the target method (which happens to
be still $a$ and $b$ in this example). 



\smalltitle{Determining Boundary Checks}
We consider a statement \code{if($c$) $s$} as a boundary check if one of the
following rules are satisfied.
We use $x$ to
denote a variable and $v$ to denote a constant.

\smallskip
\noindent Rule 1: Condition $c$ takes the form of $x==v$, $x.equals(v)$, or their
negations.

The intuition is that $s$ is special logic to compute the result for
the boundary case where $x$ equals $v$.

\smallskip
\noindent Rule 2: Condition $c$ takes the form of $x>v$, $x<v$, or their
negations, and $s$ takes the form of \code{throw $e$}.

The intuition is that the input is outside of
the input domain, and an exception should be thrown.


\subsection{Variable Ranking}
Given a target location for condition synthesis, we first try to select
variables that can be used in the expression.
Our variable ranking consists of the following steps: (1) preparing
candidate variables, (2) filtering variables by document analysis, and (3)
sorting the variables using dependency-based ordering. 
The second step is already illustrated in \secref{sec:overview}. In
the following we explain the first and the third steps. 

\smalltitle{Preparing Candidate Variables} We consider four types of
variables: (1) local variables, (2) method parameters, (3) \code{this} pointer, and (4) expressions
used in other ``if'' conditions in the current method. Strictly speaking,
the last type is not a variable, we nevertheless include it because
many defects require a complex expressions to repair
. To treat the last type unified as variables,
we assume a temporary variable is introduced for each
expression, i.e., given an expression $e$, we assume there exists a
statement \code{v=$e$} before the target conditional expression, where
\code{v} is a fresh variable.

Not all the above three types of variables can be used to synthesize a
plausible condition. If a variable
takes the same value in two test executions, but the condition are
expected to evaluate to two different values,
it is impossible to synthesize a plausible condition with the
variable. Therefore, we first filter out all such variables. The
remaining variables form the \emph{candidate variables} to be used in
condition synthesis.

\smalltitle{Sorting by Dependency}
Given a set of candidate variables, we sort them using
dependency-based ordering.

To sort the variables, we create a dependency graph between variables.
The nodes are variables, the edges are dependency relations between
variables. Concretely, we consider the following types of
intra-procedural dependencies.
\begin{itemize}
\item {\it Data Dependency}. Given an assignment statement \code{$x$=$exp$}, $x$ depends on
  all variables in $exp$.
\item {\it Control Dependency}. Given an a conditional statement such as
  \code{if($cond$) \{$stmts$\} else \{$stmts'$\}}, or
   \code{while($cond$) \{$stmts$\}},
all variables assigned in $stmts$ and $stmts'$ depend on variables in
$cond$.
\end{itemize}
Note that the dependency relations here are incomplete. For example, if
$exp$ contains a method call, extra dependencies may be caused by the
method. However,
implementing a complete dependency analysis is difficult and the
current relations are enough for ranking the variables.

\iftoggle{longversion}{
For example, \figref{fig:dependencyGraph} shows the dependency graph
of the example in \figref{fig:example1}. The dependency edges are created by the
assignment statement at lines 1 and 2.

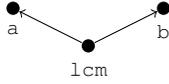
\begin{figure}[ht]
  \centering
\begin{tikzpicture}[auto,swap,->]
\tikzstyle{lib}=[circle,fill=black,minimum size=5pt,inner sep=0pt]
\node[lib] (lcm) at (1,0) [label=below:\code{lcm}]{};
\node[lib] (a) at (0,0.5) [label=below:\code{a}]{};
\draw[solid] (lcm) -> (a);
\node[lib] (b) at (2,0.5) [label=below:\code{b}]{};
\draw[solid] (lcm) -> (b);
\end{tikzpicture}
  \vspace{-0.3cm}
  \caption{\label{fig:dependencyGraph} Dependency Graph of \figref{fig:example1}}
\end{figure}

Then we perform a topological sorting to ensure the most dependent
variables are sorted first. First we identify all circles on the
graph, and collapse them as one node containing several variables.
Now the graph is acyclic and we can perform topological sorting. We first
identify the nodes with no
incoming edges (\code{lcm} in \figref{fig:dependencyGraph}), and give
them priority 0. Then we remove these variables and their outgoing
edges from the graph. Next we identify the nodes with no incoming
edges in the new graph (\code{a} and \code{b} in the example),
give them priority 1, and remove these nodes. This process stops when
there is no variable in the graph.

}{The dependency graph defines a partial order of the variables, and
  applying topological sorting on it we can obtain a total order.}
Note there may be multiple variables having the
same priority.
We further sort the variables by the distance between
the potentially faulty condition and
the assignment statement that initializes the variable. In the rest of
the paper we would use \emph{(priority) level} to refer to the priority assigned
to variables by the partial order and use \emph{rank} to refer to the rank in final total order.

To ensure the precision of the synthesized condition, we use only
 variables at the first $n$ priority levels in condition synthesis. In our
current implementation we set $n$ as 2. As will be shown later in our evaluation,
the first two levels contain a vast majority of the variables that
would be used in a conditional expression.

\subsection{Predicate Ranking}
\smalltitle{Mining Related Conditions}
We select an ordered list of predicates by predicate mining. Given a
repository of software projects, we first collect the conditional
expressions that are in a similar context to the conditional
expression being synthesized. Currently we use variable type, variable
name, and method name to determine a context. We say two variable or
method names are similar if we decompose the names by
capitalization into two sets of words, the intersection of the two
sets are not empty. We say a variable name is meaningful if its length
is longer than two. Assuming we are synthesizing a condition $c$ with
variable $x$ in the method $m$.
A conditional
expression $c'$ is considered to be in a similar context of $c$, if
(1) it contains one variable $x'$, (2) $x'$ has the same type as $x$, (3)
the name of $x'$ is similar to $x$ when the name of $x$ is meaningful, or the
name of the method surrounding $c'$ is similar to $m$
when the name of $x$ is not meaningful.

In our current implementation, we utilize the search engine of GitHub\footnote{https://github.com/search}
so that we can use all open source projects in GitHub as the
repository. Each time we invoke the search engine for returning Java
source files relating to three keywords: \code{if}, $t_x$, and $n_x$, where
$t_x$ is the type of variable $x$, and $n_x$ is the name of $x$ if the
name is meaningful, otherwise $n_x$ is the name of the surrounding
method. 


\smalltitle{Counting Predicates}
Given a conditional expression in a similar context, we extract the
predicates used in the conditional expressions. While we can extract
any predicate syntactically from the collected conditions, we choose
to consider a predefined space of predicates due to the following two
reasons. (1) As shown by an existing study~\cite{long2016analysis},
arbitrarily expanding the search space often leads to more incorrect
patches and even fewer correct patches. (2) Syntactically different
predicates may semantically be the same. For example, $>1$ and $>=2$
is semantically the same for an integer variable. If we deal with the
predicates syntactically, we may incorrectly calculate their
frequencies.

\newcommand{\ex}[1]{pred\llbracket #1 \rrbracket}
\newcommand{\ev}[1]{eval\llbracket #1 \rrbracket}
\newcommand{\m}[1]{\ensuremath{${\tt {#1}}$}}

\figref{fig:basic-predicates} shows the syntax definitions of the
predicates we considered. Basically, there are predicates comparing a
primitive variable with a constant $(==, <, >, <=, >=)$, testing equality
of an object with a constant (\code{.equals}) and testing the class of
an object (\code{instanceof}). Note operators $<=$ and $>=$ only apply to
floats. Since $x>=v$ is equivalent to $x>v-1$ for integers, we
normalize the predicates on integers by considering only $<$ and $>$.
We also normalize symmetric expressions such as $x>v$ and $v<x$ by
considering only the former.
We deliberately exclude \code{!=} operator because the synthesized
condition represents cases that are ignored by the developers, and it is
unlikely the developers ignore a large space such as \code{!=1}. More
discussion can be found in \secref{sec:discuss}.

\begin{figure}[t]
  \centering
  \[
    \begin{array}{lcl}
      preds &:=& p\_equal \mid o\_equal \mid io \mid lt \mid le \mid gt \mid ge\\
      p\_equal &:=& \m{==}\ prim\_const \\
      o\_equal &:=& \m{.equals} (obj\_const)\\
      io &:=& \m{instanceof}\ class\_name\\
      lt &:=& \m{< \ } numeric\_const \\
      le &:=& \m{<= } float\_const \\
      gt &:=& \m{> \ } numeric\_const \\
      ge &:=& \m{>= } float\_const \\
      prim\_const &:=& numeric\_const \mid boolean\_const\\
      numeric\_const &:=& integer\_const \mid float\_const
    \end{array}
  \]
\parbox{\columnwidth}{\footnotesize Here $obj\_const$ represents a constant
expression evaluating to an object (which can be determined
conservatively by static analysis), $class\_name$ represents a class
name, and $integer\_const$, $float\_const$, and $boolean\_const$ represents
constant values of integer, float, boolean type, respectively.}
\caption{\label{fig:basic-predicates} The Syntax of Predicates }
\end{figure}

We use a function $pred$ to extract a multiset of predicates from a
conditional expression. \figref{fig:basic-ex} shows part of the
definition of $pred$. We recursively traverse to the primitive predicates
and return them. \iftoggle{longversion}{Here we normalize the expressions such as
\code{$x^i$<=$v$}
and \code{$v$==$x$}. Furthermore, since a predicate $p(x)$ can be used as
$p(x)$ or $\neg p(x)$, we also consider the negation of a predicate if
it is also in our predicate space. As an example, from $x^i <= v$ we
get two predicates $\{<=v,\ >v\}$, and then we normalize the former and get
$\{<v+1, >v\}$.}{} We omit the definitions for floats and the
symmetric forms in \figref{fig:basic-ex} as they can be easily derived.

\begin{figure}[t]
  \centering
  \[
    \begin{array}{lcl}
      \ex{e_1\,\m{\&\&}\, e_2} &=& \ex{e_1} \cup \ex{e_2}\\
      \ex{e_1\, \m{||} \,e_2} &=& \ex{e_1} \cup \ex{e_2}\\
      \ex{\m{!}e} &=& \ex{e}\\
      \ex{x\m{.equals}(v)} &=& \{\m{.equals(} \ev{v}\m{)}\}\\
      \ex{x\, \m{==}\, v} &=& \{\m{==} \ev{v}\}\\
      \ex{x\ \m{instanceof}\ l} &=& \{\m{instanceof}\ l\}\\
      \ex{x^i <= v} &=& \{\m{<}\, \ev{v\m{+1}},\ \m{>}\, \ev{v}\}\\
      \ex{x^i < v} &=& \{\m{<}\, \ev{v},\ \m{>}\, \ev{v\m{-1}}\}\\
      \ex{x^i >= v} &=& \ex{x^i < v} \\
      \ex{x^i > v} &=& \ex{x^i <= v}\\
      \ex{x^f <= v} &=& \{\m{<=}\, \ev{v},\ \m{>}\, \ev{v}\}\\
      &\ldots\\
      \ex{v\, \m{==}\, x} &=& \ex{x\, \m{==} v}\\
      \ex{x^i <= v} &=& \ex{v >= x^i}\\
      &\ldots\\
      \ex{\_}  &=& \{\}
    \end{array}
  \]
\parbox{\columnwidth}{\footnotesize Here $e_1, e_2, e$ are arbitrary
  expressions, $x$ is a variable, $v$ is a constant expression, $eval$
  is a function evaluates a constant expression to a constant
  value
  ,
  $x^i$ is an integer variable, $x^f$ is float variable,
  and ``$\_$'' denotes an arbitrary expression not captured by
  previous patterns.}
  \caption{\label{fig:basic-ex} The $pred$ Function}
\end{figure}

Given a set of conditions in similar contexts, we apply the $pred$
function to each condition, and sort the predicates by their frequencies
in the conditions. To ensure the precision of our approach, we will
only consider the top $k$ predicates for condition synthesis. Currently
we set $k$ heuristically as 20. As will be
shown in the evaluation later, 20 is enough to cover the correct
predicate in most cases.

We also use a predefined set of
predicates for cases that are often ignored by developers. The current
set includes tests for Boolean true and false, such as \code{==true},
tests for minimum and maximum values, such as
\code{==Integer.MIN\_VALUE}, and tests for frequently-used JDK
interfaces, such as \code{instanceof Comparable}.



\section{Evaluation}
\subsection{Research Questions}
Our evaluation intends to answer the following research questions.
\begin{itemize}
\item RQ1: how do the three ranking techniques perform on ranking variables and predicates?
\item RQ2: how does our approach perform on real world defects?
\item RQ3: how does our approach compare with existing approaches?
\item RQ4: to what extent does 
each component of our approach contribute to the overall performance?
\end{itemize}

\subsection{Implementation 
}
We have implemented our approach as a Java program repair tool. Our
implementation is based on the source code of
Nopol
~\cite{xuan2016nopol} and the fault localization library
GZoltar
~\cite{Campos2012GZoltar}. Natural language processing is implemented
using Apache
OpenNLP
.
Our open-source implementation and the
detailed results of the experiments are
available online\footnote{\url{https://github.com/Adobee/ACS}}. 

\subsection{Data Set}
Our evaluation is performed on two datasets. The first dataset
consists of the top five most starred Java projects on GitHub as of Jul
15th, 2016. The
second dataset consists of four projects from
Defects4J~\cite{xuan2016nopol}, a bug database widely used for
evaluating Java program repair
systems~\cite{martinez2015automatic,Xuan2016History}. We use both
datasets to answer the first question and use the defects in
Defects4J to answer the rest three questions. 

\begin{table}[ht]
  \centering
  \caption{Statistics of the Defects4J Database \label{tab:subjects}}
  \begin{tabular}{|c|r|r|r|r|}
    \hline
    Project & KLoc & Test Cases & Defects & ID\\
    \hline
    \multicolumn{5}{|c|}{GitHub Projects}\\
    \hline
    Google I/O App & 62 & 269 & - & IO \\
    OkHttp & 68 & 1734 & - & Http\\
    Universal Image Loader & 13 & 13 & - & Image\\
    Retrofit & 18 & 405 & - & Retrofit\\
    Elasticsearch & 879 & 8120 & - & Search\\
    \hline
    \multicolumn{5}{|c|}{Defects4J}\\
    \hline
    JFreeChart & 146 & 2205 & 26 & Chart \\
    Apache Commons Math & 104 & 3602 & 106 & Math\\
    Joda-Time & 81 & 4130 & 27 & Time\\
    Apache Common Lang & 28 & 2245 & 65 & Lang\\
    \hline
    Total & 1388 & 22723 & 224 &-\\
    \hline
  \end{tabular}\\
\end{table}

\tabref{tab:subjects} shows basic metrics of the two datasets. Among
them, IO is an Android app for Google I/O conference. HTTP is an
efficient HTTP client. Image is an android library for image loading.
Retrofit is a type-safe HTTP client. Elasticsearch is a distributed
search engine. Chart is a library for displaying charts. Math is a
library for scientific computation. Time is a library for date/time
processing. Lang is a set of extra methods for manipulate JDK classes.

Note that Defects4J contains five projects in total. We did not use 
the fifth project, Closure, because GZoltar does not
support this project due to its customized testing format. This is consistent with an existing study~\cite{martinez2015automatic},
which also dropped Closure due to its incompatibility with GZoltar.


Since our predicate mining was
implemented on a search engine of GitHub, we may
happen to locate code pieces from the subject projects, in which the
defects were already fixed. To prevent such a bias, we used the following two
techniques: (1) our implementation automatically excludes the files from the same
project and from all known forked projects; (2) we manually reviewed the search results for all correctly
repaired defects, and deleted all results that may be a clone of the project
code.

\subsection{RQ1: Performance of the Three Techniques}
To answer the first question, we took the nine projects in our
datasets, used our three techniques to rank variables and predicates in
the conditional expressions in these projects.

\begin{table}[ht]
  \centering
  \caption{Dependency-based Ordering Performance \label{tab:ordering}}
  \begin{tabular}{|c|r|r|r|r|r|r}
    \hline
    Project & Variables & Level 1 & Level 2 & Avg.Rank & p-value\\
    \hline
    IO & 996 & 91.2\% & 7.2\% & 40.5\% & 4.3e-10\\
    Http & 773 & 92.1\% & 5.9\% & 38.1\%  & 4.5e-16\\
    Image& 246 & 89.8\% & 6.5\% &  33.7\%& 1.9e-11 \\
    Retrofit& 276 & 79.3\% & 10.9\% & 43.8\% & 1.7e-2\\
    Search& 7714 & 92.7\% & 6.1\% & 39.5\% & 5.8e-65\\
    \hline
    Chart& 3979 & 84.6\% & 9.9\% & 43.0\% & 1.2e-24\\
    Math& 2937 & 86.9\% & 9.7\% & 41.4\% & 5.1e-26\\
    Time& 1686 & 90.2\% & 7.9\% & 35.4\% & 1.2e-49\\
    Lang& 1997 & 95.5\% & 3.6\% & 40.4\%  & 7.9e-23\\
    \hline
    Total& 20604 & 89.9\% & 7.4\% & 40.3\% & 2.0e-214\\
    \hline
  \end{tabular}\\
  \parbox{\columnwidth}{\footnotesize{\smallskip ``Variables'' shows the number of
      variables in the conditional expressions. ``Level 1/2'' shows
      how many correct variables are ranked in the corresponding
      priority level. ``Avg.Rank'' shows the average rank of the correct
      variable in the normalized form. ``p-value'' shows the result of
      the wilcoxon signed-rank test. }}
\end{table}


We first evaluated dependency-based ordering. We took each conditional
expression in the subjects and checked how the variables used in the
conditions are ranked in dependency-based ordering. 
\tblref{tab:ordering} shows the results. We first consider the
priority levels of the variables. As we can see, 89.9\% of the
variables are in the first priority level, and 97.3\% of
the variables are in the first two levels, and there is no significant
difference between different types of projects. To further understand how the
correct variable is ranked in the final total order,
we normalized the
rank into [0,1] by this formula $\frac{rank -1}{total -1}$,
where $rank$ is the rank of the variable starting from 1 and $total$ is
the total number of variables. The result is shown in the column ``Avg.Rank''.
As we can see, the average rankings of all projects are significantly smaller than 50\%. We further performed a wilcoxon signed-rank test to
determine whether our ranking results are significantly different from 
random ranking, and the last column shows that the results
on all projects are significant ($< 0.05$). Further considering many of the variables may not be
repair candidates, the correct variable would be among the earliest
variables selected for synthesis.

\begin{table}[ht]
  \centering
  \caption{Document Analysis Performance \label{tab:document-analysis}}
  \begin{tabular}{|c|r|r|r|r|r|}
    \hline
    Project  & Conds & NoDoc   & NoVars  & Correct & Incorrect \\
    \hline
    IO       & 62    & 91.9\%  & 80.0\%  & 0.0\%   & 20.0\%    \\
    Http     & 294   & 99.3\%  & 100.0\% & 0.0\%   & 0.0\%     \\
    Image    & 61    & 83.6\%  & 70.0\%  & 20.0\%  & 10.0\%    \\
    Retrofit & 62    & 100.0\% & 0.0\%   & 0.0\%   & 0.0\%     \\
    Search   & 1379  & 96.9\%  & 58.1\%  & 23.3\%  & 18.6\%    \\
    Char     & 256   & 86.3\%  & 94.3\%  & 0.0\%   & 5.7\%     \\
    Math     & 774   & 36.4\%  & 85.3\%  & 10.8\%  & 3.9\%     \\
    Time     & 228   & 39.5\%  & 48.6\%  & 38.4\%  & 13.0\%    \\
    Lang     & 285   & 31.9\%  & 71.6\%  & 24.3\%  & 4.1\%     \\
    \hline
    Total    & 3401  & 73.0\%  & 75.8\%  & 18.0\%  & 6.2\%     \\
    \hline
  \end{tabular}                                                \\
  \parbox{\columnwidth}{\footnotesize{\smallskip ``Conds'' shows the number of
      exception-guarding conditions. ``NoDoc'' shows the proportion of
      conditions that do not have a JavaDoc comment. The last three
      columns show the proportions within those having a document.  ``NoVars''
      shows the proportion of conditions that have a JavaDoc comment that does
      not mention any variable. ``Correct'' shows the proportion of
      conditions where the JavaDoc comment mentions its variable. ``Incorrect'' shows the proportion of conditions
      where the set of mentioned variables by the JavaDoc comment is
      not empty but does not include the correct one.}}
\end{table}

We then evaluated document analysis. Since a comment usually
mentions only a few variables, it is clear that document analysis is effective at
filtering out variables, but it is not clear (1) whether it filters
out wrong variables, and (2) how many conditions
can benefit from document analysis. To answer the two questions, we
took all conditions guarding an exception, i.e., conditional statement
of the form \code{if ($c$) throw $X$}, and ran document analysis on
those containing one variable. 
\tblref{tab:document-analysis} shows the
result. As we can see from the table, in a vast majority of the cases (97.8\%)
the exception is undocumented, and among those documented cases,
72.6\% of the documents do not mention any variable. This result shows
that document analysis can only benefit a small number of
conditions. 
On the other hand, the false positives are significantly lower than the true positives,
indicating that the positive effect of document analysis would
outperform the negative effect.
We also manually inspected some false positives, and found
a main cause is that a word may take multiple meanings. For example, a
comment mentions a word ``value'', which actually refers to the value of
variable ``fieldType'', but in the method there happens to have a
variable ``value'', causing a false positive.

\begin{table}[ht]
  \centering
  \caption{Predicate Mining Performance \label{tab:predicate-mining}}
  \begin{tabular}{|r|r|r|r|r|r|}
    \hline
    Project  & Preds & Included & First   & wef$\leq$0 & wef$\leq$4 \\
    \hline
    IO       & 594   & 87.4\%   & 96.3\%  & 94.3\%     & 98.3\%     \\
    Http     & 433   & 80.4\%   & 95.7\%  & 90.1\%     & 97.7\%     \\
    Image    & 153   & 85.0\%   & 92.3\%  & 85.0\%     & 94.1\%     \\
    Retrofit & 187   & 54.5\%   & 100.0\% & 99.5\%     & 99.5\%     \\
    Search   & 5780  & 73.3\%   & 96.8\%  & 94.9\%     & 98.6\%     \\
    \hline
    Chart    & 3535  & 50.2\%   & 90.5\%  & 90.5\%     & 96.4\%     \\
    Math     & 1371  & 52.0\%   & 76.9\%  & 59.1\%     & 76.2\%     \\
    Time     & 1068  & 73.4\%   & 85.8\%  & 76.9\%     & 89.9\%     \\
    Lang     & 1139  & 79.4\%   & 89.8\%  & 87.4\%     & 95.6\%     \\
    \hline
    Total    & 14270 & 66.7\%   & 92.5\%  & 88.2\%     & 94.9\%     \\
    \hline
  \end{tabular}                                                     \\
  \parbox{\columnwidth}{\footnotesize \smallskip ``Preds'' shows the
    number of predicates. ``Included'' shows the percentage of
    predicates that is included in the returned list. ``First'' shows
    the percentages of predicates are ranked first among
    all predicates included. ``wef$\leq$k'' shows the percentages of cases
    where the wasted effort is smaller than or equal to $k$. }
\end{table}

Finally, we evaluated predicate mining. We first extracted the
predicates in our space from the conditional expressions in the
subject projects. For each predicate, we performed predicate mining to
retrieve a list of predicates and checked how the original predicates
were ranked. We did not use any pre-defined
predicates in this process. If there is a tie including the original
predicate, we consider the original predicate has the average rank of all predicates in the tie.
Note here many predicates cannot lead to the generation
of a patch. A necessary condition of generating patches from a
predicate $q$ is that $q$ should capture (evaluates to ``true'') the
case of the failed test execution. Since we know the original
predicate $p$ evaluates to ``true'' at the failed execution, we check
the satisfiability of $p \wedge q$, and any $q$ causing the formula
unsatisfiable cannot generate a
patch. 
We removed all such predicates that cannot generate a patch.

The result is shown in \tblref{tab:predicate-mining}. As we can see
from the table, a large proportion (66.7\%) of predicates are included
in the rank result. This confirms our assumption: the predicates are
heavily unevenly distributed. Furthermore, those included in the list
are ranked high, with a 92.5\% of the predicates are ranked first. To
further understand how often an incorrect predicate is ranked higher
than a correct one uniformly, we use the measurement ``wasted
effort''~\cite{B.Le2016}. The wasted effort is defined as the number
of incorrect predicates ranked higher than the correct one if the
correct one is included in the returned list, otherwise is defined as
the length of the returned list. As we can see from the table, the
wasted efforts is in general small: in 88.2\% of the cases there is no
wasted effort and in
94.9\% of the cases the wasted efforts are smaller
than or equal to 4.

\subsection{RQ2: Performance of \approach}
To answer this research question, we executed \approach against all the
bugs in Defects4J. Then we manually compared the generated patches with
the user patches, and deem an \approach patch to be correct only when
we can discover a sequence of semantically equivalent (basic)
transformations that turn one patch into the other. Note that this
criterion is conservative, and the reported correct patches are a subset
of all correct patches.
 Our experiments
were conducted on an Ubuntu virtual machine with i7 4790K 4.0GHz CPU
and 8G memory. We use 30 minutes as the timeout of each defect.


\begin{table}[ht]
  \centering
  \caption{Repair Results on Defects4J}  \label{tab:results}
  \begin{tabular}{|c|r|r|r|r|r|r|}
    \hline
    Project & Correct & Inc. & Precision & Recall & \parbox{0.5cm}{In Space} & \parbox{1.2cm}{Recall (In Space)}\\
    \hline
    Chart & 2 & 0 & 100.0\% & 7.7\% & 2 & 100.0\%\\
    Math & 12 & 4 & 75.0\% & 11.3\% & 12 & 100.0\%\\
    Time & 1 & 0 & 100.0\% & 3.7\% & 2 &50.0\%\\
    Lang & 2 & 2 & 50.0\% & 4.6\% & 2& 100.0\%\\
    \hline
    Total & 17 & 6 & 73.9\% & 7.6\% & 18 & 94.4\%\\
    \hline
  \end{tabular}
  \parbox{\columnwidth}{\footnotesize \smallskip ``Inc.'' stands for
    incorrect patches. There are 224 defects in our dataset.}
\end{table}

The results are shown in the first 5 columns of \tabref{tab:results}.
 Our approach generated
23 patches in total, where 17 are correct, giving a precision of
73.9\% and a recall of 7.6\%. 
The current recall
considers all defects, and many defects cannot be fixed by changing a
condition or returning an oracle, i.e., not belonging to our defect
class~\cite{Essay}.

To understand the recall within our defect class, we further manually
analyzed all Defects4J user patches and selected those that take a
form that our approach can generate. A patch is in space if (1) it
modifies an if condition, or returns a value or throws an
exception with a guarded condition, and (2) the condition contains one
variable or an expression used in other conditional statements, and
the predicate on the variable/expression is defined in
\figref{fig:basic-predicates}. 
Note that this process is also conservative, as other defects may also
be fixed by our approach using a form different from user patches. The
last two columns of \tabref{tab:results} show the results. To our
surprise, besides the 17 defects we fixed, we were only able to
further identify one defect, Time19,  whose patch is within the space of
our approach, giving our approach a recall of 94.4\%. Time19 was
not repaired because the correct variable is in the 3rd level after ranking and was filtered out.
This result
suggests that our approach is able to fix most defects in our space.

The patch generation time is short. Our approach spent in maximum 28.0 minutes
to generate a patch, with a median 5.5 minutes and a minimum 0.9
minutes. Note that since the web query time greatly depends on the
network speed, we exclude the web query time. A more elaborated
implementation could use a local repository to avoid most network
query time.



We also qualitatively reviewed the defects and the generated
patches. Our observation is that, although an \approach patch
usually takes a simple form, it can fix challenging defects. 
Our running example in \figref{fig:example1}
requires advanced knowledge with the \code{Math} library to know that
\code{abs} may return a negative value. Another example is Lang7 in method
\code{createNumber} of
class \code{NumberUtils}, which converts a string into a number. Our
approach generates the following patch for the method.
\begin{lstlisting}
+ if (str.startsWith("--"))
+    throw new NumberFormatException();
  return new BigDecimal(str);
\end{lstlisting}
The standard routine is to parse the string to the constructor of
\code{BigDecimal}, and if the string cannot be parsed, an exception
should be thrown by \code{BigDecimal}. However, though unstated in the
specification, the \code{BigDecimal}
implementation in Java JDK would accept a string with two minus sign,
but parses it into a wrong value. Thus a guard must be added. This
defect is difficult as we actually deal with a defect in JDK implementation.

\subsection{RQ3:  Comparison with Existing
  Approaches}

We compare our results with four program repairs systems,
jGenProg~\cite{martinez2015automatic},
Nopol~\cite{martinez2015automatic}, a reimplementation of
PAR~\cite{Xuan2016History} (mentioned as xPAR), and
HistoricalFix~\cite{Xuan2016History}, which are all program repair
systems that have been evaluated on Defects4J within our knowledge.
Among them, jGenProg and Nopol were evaluated on the same four projects~\cite{martinez2015automatic}
as us while
xPAR and HistoricalFix were evaluated on all the five projects~\cite{Xuan2016History}. For a
fair comparison, we took only the results on the four subjects. Please
note that in the experiments, the timeout for jGenProg and Nopol was
set to three hours~\cite{martinez2015automatic}, the timeout for xPAR
and HistoricalFix was set to 90 minutes~\cite{Xuan2016History}, and
the timeout for \approach
was 30 minutes. Though the machines for
executing the experiments were different, the results are unlikely to
favor our approach because our timeout setting is much shorter than others.

\begin{table}[ht]
  \centering
  \caption{Performance of Related Approaches}
  \label{tab:comparison}
  \begin{tabular}{|c|c|c|c|c|c|}
    \hline
    Approach & Correct & Incorrect & Precision & Recall \\
    \hline
    \approach & 17 & 6 & 73.9\% & 7.6\%  \\
    \hline
    jGenProg &  5 & 22 & 18.5\% & 2.2\% \\
    Nopol & 5 & 30 & 14.3\% & 2.2\% \\
    xPAR & 3 & --$^4$ & --$^4$ & 1.3\%$^2$ \\
    HistoricalFix$^1$ & 10(16)$^3$ & --$^4$ & --$^4$ & 4.5\%(7.1\%)$^{2,3}$\\
    \hline
  \end{tabular}\\
\parbox{\columnwidth}{\footnotesize \smallskip
$^1$The evaluation was based on manually annotated faulty methods but
not automatically located methods.\\
$^2$HistoricalFix and PAR were tested on selected 90 defects from Defects4J,
but the authors of that paper~\cite{Xuan2016History} believe all
other defects cannot be fixed.\\
$^3$HistoricalFix generated correct patches for 16 defects, but only
10 were ranked first.\\
$^4$Not reported.}
\end{table}

\tabref{tab:comparison} shows the results from different systems. From the table we can see that our approach has the highest precision
among all approaches, and is more than four times more precise than the second
precise approach. Our recall is also the highest among the five
approaches.

Some of the other state-of-the-art systems, such as
Angelix~\cite{Mechtaev} and Prophet~\cite{Long2016}, were designed for C and cannot be directly compared. Nevertheless, their
evaluations on the GenProg benchmark~\cite{EightDollar} shows that
they have a precision of 35.7\% and 38.5\%, respectively. Since our
precision is noticeably higher than them, it might be possible to combine
our approach with them to increase their precisions in future.

We also determine whether the fixed defects by \approach were also
fixed by any other approaches. We found only 2 defects were fixed by
other approaches, and 15 were fixed for the first time. This result shows that our approach can effectively
complement existing approaches on fixing more defects.

\subsection{RQ4:  Detailed Analysis of the Components}
In this section we evaluate the performance of the three ranking
techniques based on the 18 defects in search space.
\tblref{tab:performance} shows the detailed data about the repair
process of the 18 defects. From the table we analyze the performance
of the three techniques as follows.

\smalltitle{Dependency-based Ordering} As we can see from the table,
dependency-based ordering performs a significant role to select the
correct variables. As we can see from ``Can.'' column, there may be a
large number of candidate variables, up to 12 variables. After
dependency-based ordering, the correct variable is usually in the
first or the second to select, which greatly reduces the risk of
producing incorrect patches and the time for repair.

\smalltitle{Document Analysis}
Document Analysis performs a relatively small but a useful role.
From the table we can see that document analysis was only able to filter
one variable, which is consistent with the result of RQ1 as many
exceptions are undocumented. To understand how much document
analysis contributed to our result, we further reran \approach without
document analysis on all defects where
document analysis filtered variables. The result showed that document analysis
successfully prevented one incorrect patch to be generated and had not
prevented any correct patch. Therefore document analysis did contribute
to the overall performance of \approach.

\smalltitle{Predicate Mining} Predicate mining also play a significant
role of preventing incorrect fixes from being generated. In theory,
each candidate variable at a wrong location or each wrong variable
ranked higher than the correct variable at the correct location can
produce incorrect patches. This is because we can always generate
\code{$x$ == $v$} to capture the failed test case, where $x$ is the
candidate variable and $v$ is the value of $x$ in the failed test
execution. As we can see from ``Prv.'' column, many incorrect patches
have been prevented by predicate mining on the 18 defects. Further
considering the large number of defects that are not within the search
space, we can assume that predicate mining must have prevented a large
number of incorrect patches from being generated. Please note that
``Prv.'' may be larger than ``Rank'' because fault localization may
initially locate wrong locations. Furthermore, as we can see from the
``Blk.''  column, no correct patch is blocked by predicate mining.

\begin{table}[t]
  \centering
  \caption{Detailed Performance Analysis}
  \label{tab:performance}

  \begin{tabular}{|c|r|r|r|r|c|}
    \hline
    Bug ID & Can. & Flt. &  Rank & Prv. & Blk.  \\
    \hline
    \multicolumn{6}{|c|}{Fixed} \\
    \hline
    Chart14 &  2  & - & 2 & 1 & no \\
    Chart19 &  2  & - & 2 & 1 & no \\
    Math3   &  3  & - & 3 & 2 & no \\
    Math4   &  1  & - & 1 & 0 & no \\
    Math5   &  1  & - & 1 & 1 & no \\
    Math25  &  12 & 0 & 2 & 6 & no \\
    Math35  &  3  & 1 & 2 & 0 & no \\
    Math61  & 3   & - & 1 & 0 & no \\
    Math82  & 8   & 0 & 1 & 19& no \\
    Math85  &  12 & - & 1 & 2 & no \\
    Math89  &   1 & 0 & 1 & 0 & no \\
    Math90  &  2  & - & 1 & 0 & no \\
    Math93  &   1 & - & 1 & 1 & no \\
    Math99  &  3  & 0 & 1 & 0 & no \\
    Time15  &   1 & 0 & 1 & 0 & no \\
    Lang7   &  1  & 0 & 1 & 0 & no \\
    Lang24  &   3 & 0 & 1 & 2 & no \\
    \hline
    \multicolumn{6}{|c|}{Unfixed} \\
    \hline
    Time19 & 4& -&3 &3& no\\
    \hline
  \end{tabular}
\parbox{\columnwidth}{\footnotesize  \smallskip 
  ``Can.'' shows the
  number of candidate variables in the correct location. ``Flt.''
  shows how many variables were filtered out by document analysis.
  ``Ranks'' shows how the correct
  variable was ranked by
  dependency-based ordering. ``Prv.'' shows how many variables were
  prevented to generate incorrect
  patches by predicate mining. ``Blocked'' shows whether the correct
  patch was blocked by predicate mining.
}
\end{table}

Since predicate mining uses both predefined predicates and mined
predicates, we further inspect how many successful patches whose
predicates are predefined and how many are mined from GitHub. We found
that 11 (64.7\%) out of the 17 patches use mined predicates, and 6
(35.3\%) uses predefined patches. \iftoggle{longversion}{We further
  investigate whether it is possible to enlarge the set of predefined
  predicates to replace mined predicates. Among the 12 patches, we
  found three use predicate \code{==null}, which has the potential to
  be predefined. The predicates of the remaining eight patches seem to
  be difficult to be predefined.}{Many mined predicates cannot be
  easily predefined.} For example, below is the patch generated for
Math35, where the predicate tests the range of a ``rate'', which is
usually between 0 and 1. The two lines are added from two failed
tests, respectively.
\begin{lstlisting}[numbers=none,xleftmargin=0em]
+ if(elitismRate<(double)0) throw ...
+ if(elitismRate>(double)1) throw ...
\end{lstlisting}
Another example is the patch for Math5, where the predicate compares with an instance of
the \code{Complex} class.
\begin{lstlisting}[numbers=none,xleftmargin=0em]
+ if(this.equals(new Complex(0,0))) return INF;
\end{lstlisting}

\smalltitle{Templates} We also study how many defects each template
fixes. We found that 9 are fixed by exception-throwing, 6 are fixed
by value-returning, and 2 are fixed by narrowing. This result suggests
that our approach may be effective in fixing defects related to missing boundary
checks.

\section{Discussion}\label{sec:discuss}
\iftoggle{longversion}{\smalltitle{More Patches for a Defect} In this paper we focus on
generating only one patch for a defect. This is because our long-term
goal is to achieve fully automatic defect repair, and thus we need to
improve the precision of the first generated patch. On the other hand,
we may also view the generated patches as debugging aids, then we can
generate more than one patches for a defect and aim to rank the
correct patch as high as possible. The performance of our approach in
the latter scenario is a future work to be explored.}{}

\smalltitle{Alternative Method in Condition Synthesis} There are two
methods to synthesize a predicate and a variable. For example, below is
a correct patch we synthesize for Math85.
\begin{lstlisting}[]
-if (fa*fb>=0) {
+if (fa*fb>=0 && !(fa*fb==0)) {
\end{lstlisting}
In our approach, \approach mines a predicate \code{==0} to capture the
failed execution where \code{fa*fb} is zero, and then negate the
condition to get the expected result. An alternative is not to
negate conditions, and rely on predicate mining to discover a predicate
\code{!=0} that evaluates to \code{false} on failed test execution. We choose
the former approach because it gives us more control over the
predicate space to exclude the predicates that are unlikely to be
ignored by human developers. Currently we exclude \code{!=v}, as the
developers are unlikely to ignore such a large input space. If we resort to
the latter solution, we have to at least include \code{!=0} into the
predicate space, which unlikely leads to more correct patches than the
former approach but
nevertheless brings the risk of generating more incorrect patches.

\smalltitle{Generalizability} A question is whether our repair results
on Defects4J can be generalized to different types of projects. To
answer the question, we designed RQ1 that tested our ranking
techniques on different types of projects, and the results show that
dependency-based ordering and predicate mining gives a consistent
ranking among different projects. Document analysis heavily depends
on how much documentation is provided. However, as RQ4 shows, document
analysis plays a relatively minor role in affecting our results: only
one variable is removed for 18 defects. Therefore, we
believe our approach can achieve a similar precision across a wide range
of projects.







\balance
\section{Related Work}\label{sec:related}
Automatic defect repair is a hot research topic in recent years and
many different approaches have been proposed. Some recent publications~\cite{xuan2016nopol,Long2016,survey} have made
thorough survey about the field, and we discuss only the approaches that are related to condition synthesis and
patch ranking.

\smalltitle{Condition Synthesis for Program Repair}
Several program repair approaches adopt condition synthesis as part of
the repair process. Typically, a condition synthesis problem is
treated as a specialized search problem where the search space is the
all valid expressions and the goal is to pass all tests.
Nopol~\cite{xuan2016nopol} reduces the search problem as a constraint
solving problem and uses an SMT solver to solve the problem.
Semfix~\cite{SemFix} uses a similar approach to repair conditions and
other expressions uniformly. SPR~\cite{staged} uses a dedicated search
algorithm to solve the search problem. DynaMoth~\cite{durieux2016dynamoth} collects
dynamic values of primitive expressions and combines them.

Several approaches try to fix a defect by mutating the program, which may also mutate conditions. For
example, GenProg~\cite{GenProgTSE} and SearchRepair~\cite{Ke15ase} mutate programs by copying
statements from other places. PAR~\cite{PAR} mutates the
program with a set of predefined templates.

However, since the goal of these approaches is to pass all tests, it
is difficult for these approaches to achieve a high precision. As
reported~\cite{martinez2015automatic}, jGenProg (the Java
implementation of GenProg) and Nopol have a precision of 18.5\% and
14.3\% on Defects4J, respectively.

\smalltitle{Patch Ranking Techniques}
Many researchers have realized the problem of low precision and have
proposed different approaches to rank the potential patches in the
repair space, so that patches with higher probability of correctness
will be ranked higher.

DirectFix~\cite{DirectFix}, Angelix~\cite{Mechtaev} and
Qlose~\cite{dqlose} try to generate patches that make minimal changes
to the original program. DirectFix and Angelix use syntactic
difference while Qlose uses semantic differences~\cite{reps1997use}.
MintHint~\cite{MintHint} uses the statistical correlation between the
changed expression and the expected results to rank the patches.
Prophet~\cite{Long2016} learns from existing patches to prioritize
patches in the search space of SPR.
HistoricalFix~\cite{Xuan2016History} learns from existing patches to
prioritize a set of mutation operations. AutoFix~\cite{Autofix-E}
ranks the patches by the invariants used to generate
them. 

Compared with these approaches, our approach uses more refined ranking
techniques specifically designed for condition synthesis. Our approach is also
the first to utilize the locality of variables, the program document,
and the existing source code (in contrast to the existing patches) for
ranking.
\iftoggle{longversion}{

  Here we use an example to analyze why our
  approach is more fine-grained than existing approaches on ranking
  conditions. Given two plausible patches: adding a guard condition
  $a>1$ to a statement and adding a guard condition $a>2$ to the same
  statement, none of the above approaches can distinguish them.
  DirectFix and Anglix would treat the two patches as having the same
  syntactic structure and thus the same distance from the original
  program. Qlose mainly uses the coverage information and variable
  values in test execution, and the two patches would not exhibit any
  difference since they are all plausible. The case of MintHint is
  similar to Qlose as the two conditions would evaluate to the same
  value in all tests. Prophet ranks partial patches, i.e., conditions
  are reduced to only variables and the predicates are ignored, and
  the two patches are the same in the partial form. Prophet cannot
  rank full patches because the time is unaffordable. HistoricalFix
  ranks mutation operators, and it is not affordable to distinguish
  each integer as a different mutation operator. Similarly, AutoFix
  ranks patches by invariants, and it would not affordable to identify
  an invariant for each integer. On the other hand,
  our approach can still rank them by their frequencies in existing
  projects.}{The full version of
 this paper~\cite{tr} contains an example to explain why our ranking technique is
 more refined.
}


Another way to increase precision is the recently proposed
anti-patterns~\cite{tananti}. An anti-pattern blocked a class of
patches that are unlikely to be correct. Compared with anti-patterns,
our approach aims to rank all patches, including those not blocked by
anti-patterns.

DeepFix~\cite{gupta2017deepfix} uses deep learning to directly
generate patches. The precision of this approach depends on the
neutral network learned from the training set. However, so far this
approach is only evaluated on syntactic errors from students' homework
and its performance on more complex defects is yet unknown.

QACrashFix~\cite{Gao2015Fixing} is an approach that constructs
patches by reusing existing patches on StackOverflow. QACrashFix
achieves a precision of 80\%. However, this approach is limited to
crash fixes whose answers already exist on the QA site, which does
not apply to most defects fixed by our approach on Defects4J.


\smalltitle{Defect Classes with Precise Specification}
Several approaches target at defect classes where the specification is
complete, effectively avoiding the problem of weak test
suites~\cite{qi15}. Typical defect classes include memory
leaks~\cite{Gao}, where the specification is semantically equivalent
to the original program without leaks,
concurrency bugs~\cite{Deng2015,lin2014automatic,Cai16}, where the specification is
semantically equivalent to the original program without concurrency bugs,
and configuration errors~\cite{xiong2015range}, where the
specification can be interactively queried from the user. Though these
approaches have a high precision, they target totally different
defect classes compared with our work.

\smalltitle{Fix with Natural Language Processing}
Existing research has already brought natural language processing into
program repair. R2Fix~\cite{r2fix} generates patches directly from bug report
by using natural language processing to analyze bug reports. Different
from R2Fix, in our approach we utilize the document to enhance the
precision of program repair, and we still require a failed test.


\section{Conclusion}
In this paper we study refined ranking techniques for condition
synthesis, and the new program repair system achieves a relatively high precision
(73.9\%) and a reasonable recall (7.6\%) on Defects4J.  The result
indicates that studying refined ranking techniques for specific repair techniques
is promising, and calls for more studies on more different types of
repair technique.



\newpage




\bibliographystyle{IEEEtran}
\bibliography{references}

\begin{thebibliography}{10}
\providecommand{\url}[1]{#1}
\csname url@samestyle\endcsname
\providecommand{\newblock}{\relax}
\providecommand{\bibinfo}[2]{#2}
\providecommand{\BIBentrySTDinterwordspacing}{\spaceskip=0pt\relax}
\providecommand{\BIBentryALTinterwordstretchfactor}{4}
\providecommand{\BIBentryALTinterwordspacing}{\spaceskip=\fontdimen2\font plus
\BIBentryALTinterwordstretchfactor\fontdimen3\font minus
  \fontdimen4\font\relax}
\providecommand{\BIBforeignlanguage}[2]{{%
\expandafter\ifx\csname l@#1\endcsname\relax
\typeout{** WARNING: IEEEtran.bst: No hyphenation pattern has been}%
\typeout{** loaded for the language `#1'. Using the pattern for}%
\typeout{** the default language instead.}%
\else
\language=\csname l@#1\endcsname
\fi
#2}}
\providecommand{\BIBdecl}{\relax}
\BIBdecl

\bibitem{GenProg}
W.~Weimer, T.~Nguyen, C.~Le~Goues, and S.~Forrest, ``Automatically finding
  patches using genetic programming,'' in \emph{ICSE '09}, 2009, pp. 364--374.

\bibitem{PAR}
D.~Kim, J.~Nam, J.~Song, and S.~Kim, ``Automatic patch generation learned from
  human-written patches,'' in \emph{ICSE '13}, 2013, pp. 802--811.

\bibitem{Autofix-E}
Y.~Pei, C.~A. Furia, M.~Nordio, Y.~Wei, B.~Meyer, and A.~Zeller, ``Automated
  fixing of programs with contracts,'' \emph{IEEE Transactions on Software
  Engineering}, vol.~40, no.~5, pp. 427--449, 2014.

\bibitem{SemFix}
H.~D.~T. Nguyen, D.~Qi, A.~Roychoudhury, and S.~Chandra, ``Semfix: Program
  repair via semantic analysis,'' in \emph{ICSE}, 2013, pp. 772--781.

\bibitem{DirectFix}
S.~Mechtaev, J.~Yi, and A.~Roychoudhury, ``Directfix: Looking for simple
  program repairs,'' in \emph{ICSE}, 2015.

\bibitem{staged}
F.~Long and M.~Rinard, ``Staged program repair with condition synthesis,'' in
  \emph{ESEC/FSE}, 2015.

\bibitem{RSRepair}
Y.~Qi, X.~Mao, Y.~Lei, Z.~Dai, and C.~Wang, ``The strength of random search on
  automated program repair,'' in \emph{ICSE}, 2014, pp. 254--265.

\bibitem{r2fix}
C.~Liu, J.~Yang, L.~Tan, and M.~Hafiz, ``{R2Fix}: {A}utomatically generating
  bug fixes from bug reports,'' in \emph{ICST}, 2013.

\bibitem{MintHint}
S.~Kaleeswaran, V.~Tulsian, A.~Kanade, and A.~Orso, ``{MintHint}: Automated
  synthesis of repair hints,'' in \emph{ICSE}, 2014, pp. 266--276.

\bibitem{DBLP:journals/chinaf/QiMWDG12}
Y.~Qi, X.~Mao, Y.~Wen, Z.~Dai, and B.~Gu, ``More efficient automatic repair of
  large-scale programs using weak recompilation,'' \emph{{SCIENCE} {CHINA}
  Information Sciences}, vol.~55, no.~12, pp. 2785--2799, 2012.

\bibitem{qi15}
Z.~Qi, F.~Long, S.~Achour, and M.~Rinard, ``Efficient automatic patch
  generation and defect identification in {Kali},'' in \emph{ISSTA}, 2015, pp.
  257--269.

\bibitem{Tao:2014:AGP:2635868.2635873}
Y.~Tao, J.~Kim, S.~Kim, and C.~Xu, ``Automatically generated patches as
  debugging aids: A human study,'' in \emph{FSE}, 2014, pp. 64--74.

\bibitem{smith2015cure}
E.~K. Smith, E.~T. Barr, C.~Le~Goues, and Y.~Brun, ``Is the cure worse than the
  disease? overfitting in automated program repair,'' in \emph{FSE}, 2015, pp.
  532--543.

\bibitem{martinez2015automatic}
M.~Martinez, T.~Durieux, R.~Sommerard, J.~Xuan, and M.~Monperrus, ``Automatic
  repair of real bugs in java: A large-scale experiment on the {Defects4J}
  dataset,'' \emph{Empirical Software Engineering}, pp. 1--29, 2016.

\bibitem{long2016analysis}
F.~Long and M.~Rinard, ``An analysis of the search spaces for generate and
  validate patch generation systems,'' in \emph{ICSE}, 2016.

\bibitem{kitzelmann2009inductive}
E.~Kitzelmann, ``Inductive programming: A survey of program synthesis
  techniques,'' in \emph{International Workshop on Approaches and Applications
  of Inductive Programming}.\hskip 1em plus 0.5em minus 0.4em\relax Springer,
  2009, pp. 50--73.

\bibitem{Long2016}
F.~Long and M.~Rinard, ``Automatic patch generation by learning correct code,''
  in \emph{POPL}, 2016.

\bibitem{Xuan2016History}
X.-B.~D. Le, D.~Lo, and C.~{Le Goues}, ``History driven program repair,'' in
  \emph{SANER}, 2016.

\bibitem{Mechtaev}
S.~Mechtaev, J.~Yi, and A.~Roychoudhury, ``Angelix: Scalable multiline program
  patch synthesis via symbolic analysis,'' in \emph{ICSE}, 2016.

\bibitem{dqlose}
L.~D’Antoni, R.~Samanta, and R.~Singh, ``Qlose: Program repair with
  quantiative objectives,'' in \emph{CAV}, 2016.

\bibitem{EightDollar}
C.~Le~Goues, M.~Dewey-Vogt, S.~Forrest, and W.~Weimer, ``A systematic study of
  automated program repair: Fixing 55 out of 105 bugs for \$8 each,'' in
  \emph{ICSE}, 2012, pp. 3--13.

\bibitem{xuan2016nopol}
J.~Xuan, M.~Martinez, F.~Demarco, M.~Cl{\'e}ment, S.~Lamelas, T.~Durieux,
  D.~Le~Berre, and M.~Monperrus, ``Nopol: Automatic repair of conditional
  statement bugs in java programs,'' \emph{IEEE Transactions on Software
  Engineering}, 2016.

\bibitem{genesis}
F.~Long, P.~Amidon, and M.~Rinard, ``Automatic inference of code transforms and
  search spaces for automatic patch generation systems,'' MIT, Tech. Rep.
  MIT-CSAIL-TR-2016-010, 2016.

\bibitem{just2014defects4j}
R.~Just, D.~Jalali, and M.~D. Ernst, ``Defects4j: A database of existing faults
  to enable controlled testing studies for java programs,'' in \emph{ISSTA},
  2014, pp. 437--440.

\bibitem{zhai2016automatic}
J.~Zhai, J.~Huang, S.~Ma, X.~Zhang, L.~Tan, J.~Zhao, and F.~Qin, ``Automatic
  model generation from documentation for java api functions,'' in \emph{ICSE},
  2016, pp. 380--391.

\bibitem{Jones:2002:VTI:581339.581397}
J.~A. Jones, M.~J. Harrold, and J.~Stasko, ``Visualization of test information
  to assist fault localization,'' in \emph{ICSE}, 2002.

\bibitem{zhang2006locating}
X.~Zhang, N.~Gupta, and R.~Gupta, ``Locating faults through automated predicate
  switching,'' in \emph{ICSE}, 2006, pp. 272--281.

\bibitem{abreu2007accuracy}
R.~Abreu, P.~Zoeteweij, and A.~J. Van~Gemund, ``On the accuracy of
  spectrum-based fault localization,'' in \emph{TAICPART-MUTATION}, 2007, pp.
  89--98.

\bibitem{Xuan:2014:LCM:2705615.2706097}
J.~Xuan and M.~Monperrus, ``Learning to combine multiple ranking metrics for
  fault localization,'' in \emph{ICSME}, 2014, pp. 191--200.

\bibitem{steimann2013threats}
F.~Steimann, M.~Frenkel, and R.~Abreu, ``Threats to the validity and value of
  empirical assessments of the accuracy of coverage-based fault locators,'' in
  \emph{ISSTA}, 2013, pp. 314--324.

\bibitem{Xie:2013:TAR:2522920.2522924}
X.~Xie, T.~Y. Chen, F.-C. Kuo, and B.~Xu, ``A theoretical analysis of the risk
  evaluation formulas for spectrum-based fault localization,'' \emph{ACM Trans.
  Softw. Eng. Methodol.}, vol.~22, no.~4, pp. 31:1--31:40, 2013.

\bibitem{Campos2012GZoltar}
J.~Campos, A.~Riboira, A.~Perez, and R.~Abreu, ``Gzoltar: an eclipse plug-in
  for testing and debugging,'' in \emph{ASE}, 2012, pp. 378--381.

\bibitem{B.Le2016}
T.-D. {B. Le}, D.~Lo, C.~{Le Goues}, and L.~Grunske, ``{A learning-to-rank
  based fault localization approach using likely invariants},'' in
  \emph{ISSTA}, 2016, pp. 177--188.

\bibitem{Essay}
M.~Monperrus, ``A critical review of "automatic patch generation learned from
  human-written patches": Essay on the problem statement and the evaluation of
  automatic software repair,'' in \emph{ICSE}, 2014, pp. 234--242.

\bibitem{survey}
------, ``Automatic software repair: a bibliography,'' University of Lille,
  Tech. Rep. \#hal-01206501, 2015.

\bibitem{durieux2016dynamoth}
T.~Durieux and M.~Monperrus, ``Dynamoth: dynamic code synthesis for automatic
  program repair,'' in \emph{AST}, 2016, pp. 85--91.

\bibitem{GenProgTSE}
C.~Le~Goues, T.~Nguyen, S.~Forrest, and W.~Weimer, ``Genprog: A generic method
  for automatic software repair,'' \emph{Software Engineering, IEEE
  Transactions on}, vol.~38, no.~1, pp. 54--72, Jan 2012.

\bibitem{Ke15ase}
Y.~Ke, K.~T. Stolee, C.~{Le Goues}, and Y.~Brun, ``Repairing programs with
  semantic code search,'' in \emph{ASE}, 2015, pp. 295--306.

\bibitem{reps1997use}
T.~Reps, T.~Ball, M.~Das, and J.~Larus, ``The use of program profiling for
  software maintenance with applications to the year 2000 problem,'' in
  \emph{ESEC/FSE}.\hskip 1em plus 0.5em minus 0.4em\relax Springer, 1997.

\bibitem{tananti}
S.~H. Tan, H.~Yoshida, M.~R. Prasad, and A.~Roychoudhury, ``Anti-patterns in
  search-based program repair,'' in \emph{FSE}, 2016.

\bibitem{gupta2017deepfix}
R.~Gupta, S.~Pal, A.~Kanade, and S.~Shevade, ``Deepfix: Fixing common c
  language errors by deep learning,'' in \emph{AAAI}, 2017.

\bibitem{Gao2015Fixing}
Q.~Gao, H.~Zhang, J.~Wang, and Y.~Xiong, ``Fixing recurring crash bugs via
  analyzing q\&a sites,'' in \emph{ASE}, 2015, pp. 307--318.

\bibitem{Gao}
Q.~Gao, Y.~Xiong, Y.~Mi, L.~Zhang, W.~Yang, Z.~Zhou, B.~Xie, and H.~Mei, ``Safe
  memory-leak fixing for c programs,'' in \emph{ICSE}, 2015.

\bibitem{Deng2015}
D.~Deng, G.~Jin, M.~de~Kruijf, A.~Li, B.~Liblit, S.~Lu, S.~Qi, J.~Ren,
  K.~Sankaralingam, L.~Song, Y.~Wu, M.~Zhang, W.~Zhang, and W.~Zheng, ``Fixing,
  preventing, and recovering from concurrency bugs,'' \emph{Science China
  Information Sciences}, vol.~58, no.~5, pp. 1--18, 2015.

\bibitem{lin2014automatic}
Y.~Lin and S.~S. Kulkarni, ``Automatic repair for multi-threaded programs with
  deadlock/livelock using maximum satisfiability,'' in \emph{ISSTA}, 2014, pp.
  237--247.

\bibitem{Cai16}
Y.~Cai and L.~Cao, ``Fixing deadlocks via lock pre-acquisitions,'' in
  \emph{ICSE}, 2016.

\bibitem{xiong2015range}
Y.~Xiong, H.~Zhang, A.~Hubaux, S.~She, J.~Wang, and K.~Czarnecki, ``Range
  fixes: Interactive error resolution for software configuration,''
  \emph{Software Engineering, IEEE Transactions on}, vol.~41, no.~6, pp.
  603--619, 2015.

\end{thebibliography}
%

\end{document}